\documentclass[epj]{webofc}\usepackage[varg]{txfonts}\usepackage{bm}
\usepackage{color}\usepackage{amsmath}\usepackage{slashed}
\woctitle{QCD@Work 2016}\definecolor{violet}{rgb}{0.4336,0,1}
\begin{document}\title{Janus-Facedness of the Pion:\\Analytic
Instantaneous Bethe--Salpeter Models}\author{Wolfgang
Lucha\inst{1}\fnsep\thanks{\email{Wolfgang.Lucha@oeaw.ac.at}}}
\institute{Institute for High Energy Physics, Austrian Academy of
Sciences, Nikolsdorfergasse 18, A-1050 Vienna, Austria}

\abstract{Inversion enables the construction of interaction
potentials underlying --- under fortunate circumstances even
analytic --- instantaneous Bethe--Salpeter descriptions of all
lightest pseudoscalar mesons as quark--antiquark bound states of
Goldstone-boson nature.}\maketitle

\section{Introduction: quark--antiquark bound states of
Goldstone-boson identity}\label{Sec:GQABS}Within quantum
chromodynamics, the pions or, as a matter of fact, all \emph{light
pseudoscalar mesons\/} must be interpretable as both
\emph{quark--antiquark bound states\/} and almost massless
(pseudo) \emph{Goldstone~bosons\/} related to the spontaneously
(and, to a minor extent, also explicitly) broken chiral symmetries
of QCD.

Relativistic quantum field theory describes bound states by their
Bethe--Salpeter amplitudes, $\Phi(p),$ controlled by the
\emph{homogeneous Bethe--Salpeter equation\/} defined (for two
bound particles of individual and relative momenta $p_{1,2}$ and
$p$) by their full propagators $S_{1,2}(p_{1,2})$ and the integral
kernel~$K(p,q)$~that encompasses their interactions (notationally
suppressing dependences on the total momentum~$p_1+p_2$):$$\Phi(p)
=\frac{\rm i}{(2\pi)^4}\,S_1(p_1)\int{\rm
d}^4q\,K(p,q)\,\Phi(q)\,S_2(-p_2)\ .$$

The application of suitably adapted \emph{inversion\/} techniques
\cite{WL13} allows us to retrieve all the underlying interactions
--- rooted, of course, in QCD --- analytically in the form of a
(configuration-space) \emph{central potential\/} $V(r),$
$r\equiv|\bm{x}|,$ from presumed solutions to the Bethe--Salpeter
equation \cite{WLPoS}. By that, we are put in a position to
construct \emph{exact analytic\/} Bethe--Salpeter
\emph{solutions\/} for all massless pseudoscalar mesons
\cite{WL15} in the sense of establishing in a rigorous manner the
analytic relationships between interactions and resulting
solutions: all analytic findings \cite{WL16} can be confronted
with associated numerical outcomes~\cite{WL16n}.

\section{Sequence of simplifying assumptions crucial for the
inversion formalism}\label{Sec:CSAS}By a few steps, we cast the
Bethe--Salpeter equation into a shape that allows us to talk about
potentials.

\begin{enumerate}\item Assuming, for each involved quark, both
\emph{instantaneous\/} interactions and \emph{free\/} propagation,
with a mass dubbed as \emph{constituent\/}, simplifies the
Bethe--Salpeter equation to a bound-state equation~for the
\emph{Salpeter amplitude\/} $\phi(\bm{p}),$ obtained from the
Bethe--Salpeter amplitude by integration over~$p_0$:
$$\phi(\bm{p})\propto\int{\rm d}p_0\,\Phi(p)\ .$$Generically, for
a spin-$\frac{1}{2}$ fermion and a spin-$\frac{1}{2}$ antifermion
of equal constituent masses $m,$ bound to a spin-singlet state
(which, for instance, clearly is the case for any such
pseudoscalar state),~its three-dimensional wave function involves
just two independent components, here called
$\varphi_{1,2}(\bm{p})$:$$\phi(\bm{p})=\left[\varphi_1(\bm{p})\,
\frac{\gamma_0\,(\bm{\gamma}\cdot\bm{p}+m)}{E(p)}
+\varphi_2(\bm{p})\right]\gamma_5\ ,\qquad
E(p)\equiv\sqrt{\bm{p}^2+m^2}\ ,\qquad p\equiv|\bm{p}|\ .$$\item
Upon supposing that the quark interactions in the kernel respect
spherical and Fierz symmetries, our bound-state equation for
$\phi(\bm{p})$ collapses to the system of \emph{coupled radial
eigenvalue equations\/}
$$2\,E(p)\,\varphi_2(p)+2\int\limits_0^\infty\frac{{\rm
d}q\,q^2}{(2\pi)^2}\,V(p,q)\,\varphi_2(q)=\widehat
M\,\varphi_1(p)\ ,\qquad 2\,E(p)\,\varphi_1(p)=\widehat
M\,\varphi_2(p)\ ,\qquad q\equiv|\bm{q}|\ ,$$for the bound-state
mass eigenvalue $\widehat M$ \cite{WL07}. Therein, $V(r)$ enters
via its Fourier--Bessel transform$$V(p,q)\equiv\frac{8\pi}{p\,q}
\int\limits_0^\infty{\rm d}r\sin(p\,r)\sin(q\,r)\,V(r)\ .$$\item
In the strictly massless (Goldstone) case $\widehat M=0,$ the
system decouples: one Salpeter component, $\varphi_1(p),$ is
doomed to vanish, $\varphi_1(p)\equiv0,$ whereas the surviving
Salpeter component $\varphi_2(p)$ satisfies $$E(p)\,\varphi_2(p)
+\int\limits_0^\infty\frac{{\rm d}q\,q^2}{(2\pi)^2}\,V(p,q)\,
\varphi_2(q)=0\ .$$Denoting the Fourier--Bessel transform of the
kinetic term $E(p)\,\varphi_2(p)$ by $T(r),$ the potential~$V(r)$
may be simply read off from the configuration-space representation
of this bound-state equation:$$T(r)+V(r)\,\varphi_2(r)=0\qquad
\Longrightarrow\qquad V(r)=-\frac{T(r)}{\varphi_2(r)}\
.$$\end{enumerate}

\section{Constraints on lightest-pseudoscalar-meson Bethe--Salpeter
amplitudes}\label{Sec:CBSA}Information on the input Salpeter
component $\varphi_2(p)$ can be gained from the \emph{full quark
propagator\/} $S(p),$ which is determined by its mass function
$M(p^2)$ and its wave-function renormalization function~$Z(p^2)$:
$$S(p)=\frac{{\rm i}\,Z(p^2)}{\slashed{p}-M(p^2)+{\rm
i}\,\varepsilon}\ ,\qquad\slashed{p}\equiv p^\mu\,\gamma_\mu\
,\qquad\varepsilon\downarrow0\ .$$Studies of $S(p)$ within the
Dyson--Schwinger framework, preferably done in Euclidean space
signalled by \underline{underlined} quantities, allow for pivotal
insights. \emph{In the chiral limit\/}, a Ward--Takahashi identity
links \cite{PM97a} this quark propagator to the flavour-nonsinglet
\emph{pseudoscalar-meson Bethe--Salpeter amplitude\/}
\cite{WL15}:$$\displaystyle\Phi(\underline{k})\approx
\frac{M(\underline{k}^2)}{\underline{k}^2+M^2(\underline{k}^2)}\,
\underline{\gamma}_5+\mbox{subleading contributions}\ .$$First, in
order to devise \emph{analytically\/} accessible scenarios, we
exploit two crucial pieces of information:\begin{enumerate}\item
\emph{In the chiral limit\/}, phenomenologically sound
Dyson--Schwinger studies \cite{PM97b} imply, for the quark mass
function $M(\underline{k}^2),$ at large Euclidean momenta
$\underline{k}^2$ a decrease essentially proportional to
$1/\underline{k}^2.$\item From axiomatic quantum field theory, we
may deduce \cite{CDR08} that the presence of an \emph{inflection
point at finite space-like momenta\/} $\underline{k}^2>0$ in the
quark mass function $M(\underline{k}^2)$ entails
\emph{colour~confinement\/}.
\end{enumerate}Of course, any imposition of such kind of
requirements on $M(\underline{k}^2)$ has to be \emph{reflected
by\/} $\Phi(\underline{k}).$ An \emph{ansatz\/} for
$\Phi(\underline{k})$ compatible with both constraints, involving
a mass parameter, $\mu,$ and a mixing parameter, $\eta,$~is
$$\Phi(\underline{k})=\left[\frac{1}{(\underline{k}^2+\mu^2)^2}
+\frac{\eta\,\underline{k}^2}{(\underline{k}^2+\mu^2)^3}\right]
\underline{\gamma}_5\ ,\qquad\mu>0\ ,\qquad\eta\in\mathbb{R}\ .$$
An integration of this $\Phi(\underline{k})$ with respect to the
time component of the Euclidean momentum $\underline{k}$
results~in$$\varphi_2(p)\propto\frac{1}{(p^2+\mu^2)^{3/2}}+\eta\,
\frac{p^2+\mu^2/4}{(p^2+\mu^2)^{5/2}}\ ,\qquad p\equiv|\bm{p}|\
,$$in configuration space expressible in terms of modified Bessel
functions of the second kind $K_\sigma(z)$~\cite{AS}:
\begin{equation}\varphi_2(r)\propto4\,(1+\eta)\,K_0(\mu\,r)
-\eta\,\mu\,r\,K_1(\mu\,r)\ .\label{Eq:phir}\end{equation}For
$\eta$ values satisfying $\eta<-1$ or $\eta>0,$ $\varphi_2(r)$ has
one zero, which clearly induces a \emph{singularity\/}~in~$V(r).$

\section{Analytic outcomes \cite{WL15,WL16} for interquark
potentials exhibiting confinement}\label{Sec:CPAR}For a few
particular values of the dimensionless ratio $m/\mu,$ the
\emph{analytic\/} expression of $V(r)$ can be found
\cite{WL15,WL16}. (Throughout this section, any quantity has to be
understood in units of the adequate~power of $\mu.$) As a
consequence of our ansatz for $\Phi(\underline{k}),$ giving rise
to the particular form (\ref{Eq:phir}) of $\varphi_2(r),$ for
$\eta\ne-1$ each extracted $V(r)$ will develop, at the spatial
origin $r=0,$ a logarithmically softened Coulomb~singularity:
$$V(r)\xrightarrow[r\to0]{}\frac{\mbox{const}}{r\ln r}
\xrightarrow[r\to0]{}-\infty\qquad(\mbox{const}>0)\qquad\mbox{for}\
\eta\ne-1\ .$$

\subsection{\boldmath Analytically manageable scenario of massless
quarks, i.e., of constituent mass $m=0$}\label{Subsec:AMSMQ}
\begin{figure}[hbt]\centering\includegraphics[scale=1.20296,clip]
{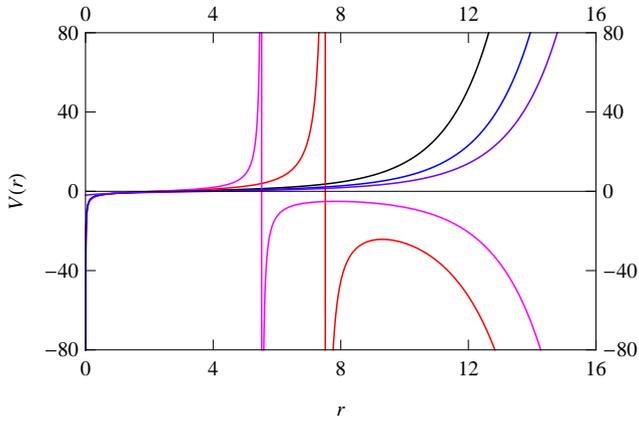}\caption{Configuration-space interquark
potential $V(r)$ of the Fierz-symmetric kernel $K(p,q),$ for the
constituent quark mass $m=0$ and mixture $\eta=0$ \cite{WL15}
(black), $\eta=1$ (\textcolor{red}{red}), $\eta=2$
(\textcolor{magenta}{magenta}), $\eta=-0.5$
(\textcolor{blue}{blue}), or
$\eta=-1$~(\textcolor{violet}{violet}).}\label{Fig:ARPm0}\end{figure}

For
our choice of $\varphi_2(r),$ $V(r)$ involves both modified Bessel
($I_n$) and Struve (${\bf L}_n$) functions \cite{AS}
($n\in\mathbb{N}$), and rises in a confinement-betraying manner to
infinity either at the zero of $\varphi_2(r)$ or for $r\to\infty$
(Fig.~\ref{Fig:ARPm0}):$$V(r)=\frac{\pi\,[4+\eta\,(4+r^2)]\,[{\bf
L}_0(r)-I_0(r)]+\pi\,(4+5\,\eta)\,r\,[{\bf L}_1(r)-I_1(r)]+4\,
(2+3\,\eta)\,r}{2\,r\,[4\,(1+\eta)\,K_0(r)-\eta\,r\,K_1(r)]}\ .$$

\subsection{\boldmath Analytically expressible observation for quarks
with common constituent mass $m=\mu$}\label{Subsec:AECQCM}
\begin{figure}[hbt]\centering\includegraphics[scale=1.20296,clip]
{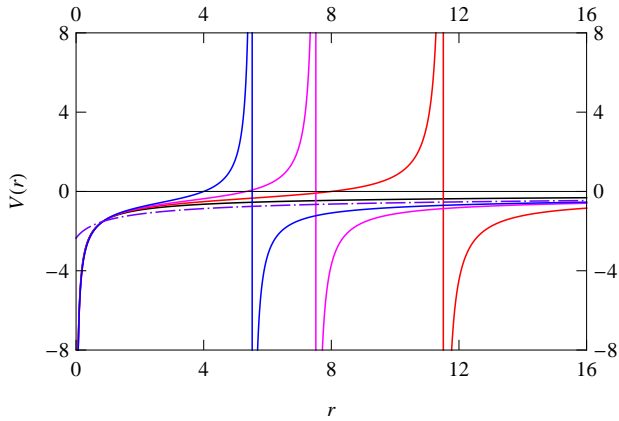}\caption{Configuration-space interquark
potential $V(r)$ of the Fierz-symmetric kernel $K(p,q),$ for the
constituent quark mass $m=1$ and mixture $\eta=0$ \cite{WL15}
(black), $\eta=0.5$ (\textcolor{red}{red}), $\eta=1$
(\textcolor{magenta}{magenta}), $\eta=2$ (\textcolor{blue}{blue}),
and $\eta=-1$~(\textcolor{violet}{violet}).}\label{Fig:ARPmu}
\end{figure}

For
$m=\mu,$ the kinetic term $T(r)$ is a mixture of Yukawa and
exponential behaviour, whence (cf.~Fig.~\ref{Fig:ARPmu})
$$V(r)=-\frac{\pi\,[8+\eta\,(8-3\,r)]\exp(-r)}
{4\,r\,[4\,(1+\eta)\,K_0(r)-\eta\,r\,K_1(r)]}\xrightarrow[r\to\infty]{}
-\frac{\mbox{const}}{\sqrt{r}}\xrightarrow[r\to\infty]{}0\qquad
(\mbox{const}>0)\ .$$

\section{Reliability check of findings: numerical determination of
the potential \cite{WL16n}}\label{Sec:RND}
\begin{figure}[hbt]\centering\begin{tabular}{cc}
\includegraphics[scale=1.2139,clip]{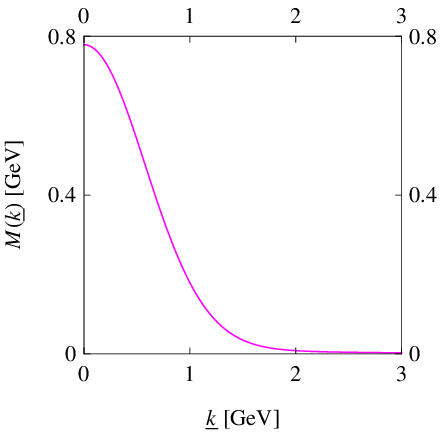}&
\includegraphics[scale=1.20296,clip]{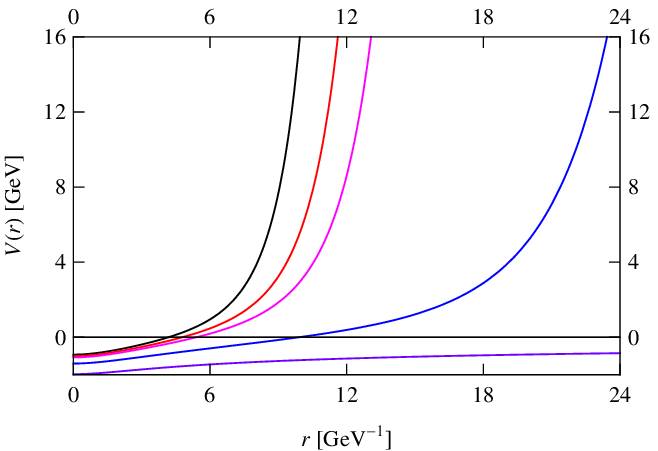}\\(a)&(b)
\end{tabular}\caption{(a) Mass function $M(\underline{k})$ deduced
from the Dyson--Schwinger model of Ref.~\cite{PM97b} for the quark
propagator. (b) Configuration-space interquark potential $V(r)$
numerically determined from $M(\underline{k}^2),$ for
constituent~quark~mass $m=0$ (black), $m=0.35\;\mbox{GeV}$
(\textcolor{red}{red}), $m=0.5\;\mbox{GeV}$
(\textcolor{magenta}{magenta}), $m=1.0\;\mbox{GeV}$
(\textcolor{blue}{blue}), and $m=1.69\;\mbox{GeV}$
(\textcolor{violet}{violet})~\cite{WL16n}.}\label{Fig:DSE-MV}
\end{figure}

Our findings may be
scrutinized by use of the chiral-limit quark mass function's
pointwise form~$M(\underline{k}^2),$ provided graphically in
Ref.~\cite{PM97b} and shown in Fig.~\ref{Fig:DSE-MV} as
$M(\underline{k})$ with
$\underline{k}\equiv(\underline{k}^2)^{1/2},$ which we
parametrize~by $$M(\underline{k})=0.708\;\mbox{GeV}
\exp\left(-\frac{\underline{k}^2}{0.655\;\mbox{GeV}^2}\right)
+\frac{0.0706\;\mbox{GeV}}{\left[1+\left(\frac{\underline{k}^2}
{0.487\;{\rm GeV}^2}\right)^{1.48}\right]^{0.752}}\ .$$Note that
the \emph{product\/} of the two exponents in the second term above
yields $1.48\times0.752\approx1.1,$ which is pretty close to
unity, as demanded by the large-$\underline{k}$ constraint.
Feeding this $M(\underline{k})$ parametrization into our inversion
procedure, we obtain potentials that are finite at $r=0$ and, for
sufficiently small $m,$ rise with $r$ to infinity but, for large
$m,$ remain negative, as illustrated in Fig.~\ref{Fig:DSE-MV} for
selected constituent~mass~values.

\section{Summary of results, observations, discussion, conclusion,
perspectives}\label{Sec:SRODCP}We constructed confining potentials
$V(r)$ that in cooperation with a Fierz-symmetric interaction
kernel describe massless pseudoscalar quark--antiquark bound-state
solutions of the Bethe--Salpeter equation. This is possible even
analytically if focusing to specific aspects of the quark mass
function's behaviour. Two obstacles call for a particularly
careful treatment: Numerically, $M(p^2)$ is known for only a
limited range of $p^2.$ For large $r,$ both $T(r)$ and
$\varphi_2(r)$ approach zero; thus, pinning down $V(r)$ in the
limit $r\to\infty$ boils down to a division of zero by zero.
Dropping the free quark propagation constraint \cite{WL05:LS}
allows us to thoroughly take into account the effects of $M(p^2)$
\emph{and\/} the quark wave-function renormalization~\cite{WL16i}.

\end{document}